\documentclass[letterpaper,twocolumn]{jpsj3_arxiv} 

\usepackage[dvipdfmx]{graphicx}
\usepackage[dvipdfmx]{color}

\usepackage{braket}

\usepackage{cite}

\usepackage{color}

\usepackage{comment}

\begin{document}

\title{Excitonic Correlations, Spin-State Ordering, and Magnetic-Field Effects in One-Dimensional Two-Orbital Hubbard Model for Spin-Crossover Region}
\author{Koya Kitagawa$^1$ and Hiroaki Matsueda$^{2,3,4}$
}
\inst{
$^1$Department of Physics, Graduate School of Science, Tohoku  University\\
$^2$Department of Applied Physics, Graduate School of Engineering, Tohoku University,\\
$^3$Center for Spintronics Research Network, Tohoku University,\\
$^4$Center for Spintronics Integrated Systems, Tohoku University
}
\recdate{\today}

\abst{
The electronic properties of excitonic insulators have been examined precisely in recent years. Pictures of exciton condensation may be applied to the spin-state transition observed in perovskite cobalt oxides. We examine the crystal-field and magnetic-field dependences of spatial spin structures on the basis of the density matrix renormalization group method using an effective model for the one-dimensional two-orbital Hubbard model in strong-coupling limit. We find an excitonic insulating (EI) phase and a spin-state ordering (SSO) phase in the intermediate region between low-spin and high-spin phases.
In the EI phase, spin-triplet excitons are spatially fluctuating due to quantum effects, and an incommensurate spin correlation realizes. 
The analyses of a spin gap and degeneracy of entanglement spectra suggest the realization of the Haldane-like edge state in the EI phase.
In the SSO phase, 3-fold or incommensurate SSO structures realize depending on the crystal-field splitting. These structures are stabilized as a result of the competition of exchange interactions between spin states. 
}
\maketitle

\section{Introduction}
The spin and orbital degrees of freedom provide a variety of physics in strongly correlated electron systems such as transition metal compounds\cite{RevModPhys.70.1039}. Fe or Co ions in such compounds change their spin states in response to temperature and pressure. For these ions, the competition between the crystal-field splitting and the Hund coupling causes the spin-state transition in the $3d$ orbital. LaCoO$_3$ is a typical perovskite cobalt oxide that shows a spin-state transition. The magnetic susceptibility and electrical conductivity of LaCoO$_3$ increase rapidly as temprature increases below 100 K \cite{PhysRevB.53.R2926, JPSJ.67.290}.

These temperature dependences are explained by three competing $3d$ electron configurations of the Co$^{3+}$ ion, which are a low-spin state with $s=0$ for $(t_{2g})^6(e_g)^0$, an intermediate-spin state with $s=1$ for $(t_{2g})^5(e_g)^1$, and a high-spin state with $s=2$ for $(t_{2g})^4(e_g)^2$. The competition between these spin states also causes exotic phenomena such as giant magnetoresistance \cite{PhysRevB.54.16044}, magnetic clustering \cite{JPSJ.78.093702, PhysRevB.80.052402},  and ferroelectricity\cite{ja102987d}.\par

Recently, the possibility of realizing excitonic insulators has been suggested in both experimental and theoretical studies.
\cite{
PhysRevB.89.115134,
PhysRevB.90.235112,
PhysRevB.93.205136,
JPSJ.85.083706,
PhysRevB.97.155114,
PhysRevLett.103.026402,
PhysRevB.90.155116
}
In excitonic insulators, a macroscopic number of electron-hole pairs are spontaneously generated by Coulomb interaction, and the pairs move coherently. This is called exciton condensation. 
In the vicinity of the spin-state transition, the fluctuation between competing spin states results in the excitonic insulating (EI) state.
Therefore, considering the EI phase is important around the spin-state transition.\par

In the perovskite cobalt oxide Pr$_{0.5}$Ca$_{0.5}$CoO$_3$, a first-order metal-insulator phase transition around 90K has attracted attention in regards to the excitonic insulator.
\cite{j.elspec.2005.01.225, PhysRevB.90.235112}
A first-order phase transition observed in high-magnetic-field experiments on LaCoO$_3$ has also been discussed in terms of its spin-state transition and exciton condensation.
\cite{
PhysRevLett.109.037201,
PhysRevB.93.220401,
PhysRevLett.125.177202,
srep30510
}
Besides the EI phase, a spin-state ordering (SSO) phase is another candidate for the magnetically induced phase in LaCoO$_3$.
In the SSO phase, the spin state is crystallized and shows a superlattice structure. Some spatial patterns have been suggested in numerical analyses.
\cite{PhysRevLett.109.037201,JPSJ.85.083706}.
The SSO phase has been observed in LaCoO$_3$ thin films on substrates [(LaAlO$_{3}$)$_{0.3}$(SrAl$_{0.5}$Ta$_{0.5}$O$_3$)$_{0.7}$], where various superlattice structures of spin states have been observed.
\cite{PhysRevLett.111.027206,PhysRevB.92.195115,JPSJ.85.023704}
In such thin-film experiments, we can vary the spin state by controlling the crystal-field splitting, which is determined by the lattice constant of the substrate.\par

As described above, in the spin-state transition, the EI and SSO phases can be realized due to the spin and orbital degrees of freedom of $3d$ electrons. To understand the physics in the spin-state transition comprehensively, we need to develop a theory of multi-orbital strongly correlated electron systems that takes into account various types of spin structures and exciton condensation.\par
In this study, we examine the realization of the EI phase and various types of SSO on the basis of numerical analysis of the two-orbital Hubbard model (TOHM), which is the simplest strongly correlated model that describes spin-state transition.
In previous theoretical studies, the realization of EI and SSO phases in the two-dimensional (2D) TOHM was confirmed under the framework of mean-field approximation and DMFT.\cite{0953-8984/27/33/333201, JPSJ.85.083706, PhysRevB.97.155114}
However, in analyses using these methods, the order parameters or the spatial structures are assumed.
In this study, the density matrix renormalization group method (DMRG)
\cite{PhysRevLett.69.2863, PhysRevB.48.10345} is performed on a low-energy effective model of the one-dimensional (1D) TOHM.
With DMRG, we can perform an accurate analysis with a large system size without any assumptions on the various types of orderings or their spatial structures. Although the dimensionality is limited to 1D due to the limitation of the DMRG, an accurate picture of a 1D quantum system can be a probe to understand 2D or 3D ones.
The main goal of this study is to make a ground-state phase diagram for the 1D TOHM in the vicinity of spin-state transition and to reveal the spatial structure in each phase.
\par
In the analysis of the ground state of the low-energy effective model of TOHM, we find a EI phase and a SSO phase in the intermediate region between the low-spin (LS) phase and the high-spin (HS) phase. 
In the EI phase, an incommensurate spin correlation occurs, and its characteristic wavenumber changes continuously depending on the crystal field splitting.
In the SSO phase, various types of SSO structures are found depending on the crystal-field splitting.
As a particularly stable SSO structure, the 3-fold structure of LS/HS/HS is stabilized as a result of competing interactions between the nearest spin states.
In the analyses of magnetic-field effects, we find the existence of a Haldane gap not only in the HS phase but also in the EI phase.
In the magnetization process in the LS phase, a magnetized EI phase is found.
In the magnetization process in the EI phase, we find that the magnetization process splits into two different processes. 
\par
This paper is organized as follows.
In Sect. \ref{sec:model_and_method}, we derive a low-energy effective model from the TOHM and describe the typical properties of the model. In Sect. \ref{sec:results}, we show the results of DMRG analyses: 
a ground-state phase diagram in SubSect. \ref{sec:results}.1,
spatial structures of the EI and SSO phases in SubSect. \ref{sec:results}.2, 
a phase diagram in a magnetic field  in SubSect. \ref{sec:results}.3,
the property of the entanglement entropy in each phase in SubSect. \ref{sec:results}.4,
and the degeneracy of the entanglement spectra in SubSect. \ref{sec:results}.5.
In Sect. \ref{sec:summary}, we summarize the paper.

\section{Model and Method}
\label{sec:model_and_method}
We start with TOHM, which is defined as
\begin{align}
\mathcal{H_{\mathrm{TOHM}}}
=&
-\sum_{\langle ij \rangle \lambda \sigma}
t_{\lambda}
\left(
c_{i\lambda\sigma}^\dagger
c_{j\lambda\sigma}
+\mathrm{H.c.}
\right)
+
\Delta\sum_{i\sigma}n_{ia\sigma}
\notag
\\
&+
U\sum_{i\lambda}n_{i\lambda\uparrow}n_{i\lambda\downarrow}
+
U'\sum_{i\sigma}n_{ia\sigma}n_{ib\sigma}
\notag
\\
&+
J\sum_{i\sigma\sigma'}
c_{ia\sigma}^\dagger
c_{ib\sigma'}^\dagger
c_{ia\sigma'}
c_{ib\sigma}
\notag
\\
&+I\sum_{i\lambda\neq\lambda'}
c_{i\lambda\uparrow}^\dagger
c_{i\lambda\downarrow}^\dagger
c_{i\lambda'\downarrow}
c_{i\lambda'\uparrow}.
\label{eq:tohm}
\end{align}
We define the annihilating operator of an electron at site $i$, orbital $\lambda(=a,b)$, and spin $\sigma(=\uparrow,\downarrow)$ as $c_{i\lambda\sigma}$. The electron number operator  is defined as $n_{i\lambda\sigma}=c_{i\lambda\sigma}^\dagger c_{i\lambda\sigma}$.
The first term in Eq. (\ref{eq:tohm}) describes the electron hopping between the same orbitals at the nearest sites $\braket{ij}$ with its amplitude $t_\lambda$. The second term describes the crystal-field splitting with its amplitude $\Delta$. The remaining terms represent the onsite Coulomb interactions, where $U$ is the intra-orbital Coulomb interaction, $U'$ is the inter-orbital Coulomb interaction, $J$ is the Hund coupling, and $I$ is the pair hopping. We use the relational equations for the Coulomb interactions in an isolated ion: $U=U'+2J$ and $I=J$. We focus on the case in which two electrons exist per site on average.
\par
In the case of the spin-state transition, where $t_\lambda\ll U, U', J, \Delta$,
the low-energy effective Hamiltonian of Eq. (\ref{eq:tohm}) is derived by Kanamori et al.\cite{PhysRevLett.107.167403} by treating the electron hopping terms as perturbations and truncating high-energy bases.
The local bases of the effective Hamiltonian are

\begin{align}
\ket{L}&=
\left(
f
c_{b\uparrow}^\dagger 
c_{b\downarrow}^\dagger 
-
g
c_{a\uparrow}^\dagger 
c_{a\downarrow}^\dagger 
\right)
\ket{0},
\label{eq:ls_state}
\\
\ket{H_{+1}}
&=
c_{a\uparrow}^\dagger 
c_{b\uparrow}^\dagger 
\ket{0},
\\
\ket{H_{0}}
&=\frac{1}{\sqrt{2}}
\left(
c_{a\uparrow}^\dagger
c_{b\downarrow}^\dagger 
+
c_{a\downarrow}^\dagger 
c_{b\uparrow}^\dagger 
\right)
\ket{0},
\\
\ket{H_{-1}}
&=
c_{a\downarrow}^\dagger
c_{b\downarrow}^\dagger 
\ket{0},
\label{eq:hm_state}
\end{align}
where $\ket{0}$ is a vacuum. The factors in Eq. (\ref{eq:ls_state}) are given as $f=1/\sqrt{1+(\Delta'-\Delta)^2/I^2}$ and $g=\sqrt{1-f^2}$, where $\Delta'=\sqrt{\Delta^2+I^2}$.
$\ket{L}$ is the LS state, where the amplitude of the one-site spin is $s=0$.
In the case of $I=0$, then $f=1$ and $g=0$, and $\ket{L}$ is the state in which two electrons occupy the low-energy orbital.
If $I$ becomes finite, then $\ket{L}$ contains the state in which two electrons occupy the high-energy orbital as hybridization.
$\ket{H_{s^z=0,\pm 1}}$ are HS states, where the amplitude of the one-site spin is $s=1$.
In this paper, the terms "LS state" and "HS state" refer to the one-site bases and not refer to global ones.
\par
We truncate the following high-energy bases which has two electrons in a site,
\begin{align}
\ket{uL}&=
\left(
g
c_{b\uparrow}^\dagger 
c_{b\downarrow}^\dagger 
+
f
c_{a\uparrow}^\dagger 
c_{a\downarrow}^\dagger 
\right)
\ket{0},
\label{eq:uls_state}
\\
\ket{iL}
&=\frac{1}{\sqrt{2}}
\left(
c_{a\uparrow}^\dagger
c_{b\downarrow}^\dagger 
-
c_{a\downarrow}^\dagger 
c_{b\uparrow}^\dagger 
\right)
\ket{0},
\label{eq:ils_state}
\end{align}
where $\ket{uL}$ is the state in which two electrons mainly occupies the high energy orbital,
and $\ket{iL}$ is the state in which one electron in the high-energy orbital and the other electron in the low-energy orbital form a singlet. This state is regarded as the singlet-exciton state while $\ket{H_{0,\pm 1}}$ are regarded as the triplet-exciton states.
If $J>0$, triplet-exciton states become more stable than the singlet one. This behaviour has been presented in previous study\cite{PhysRevB.90.245144}.
In this study, we focus on the spin-state transition, where $\Delta$ and $J$ are large. In this region, $\ket{uL}$ and $\ket{iL}$ are negligible.
\par
The second order perturbation of $t_a$ and $t_b$ on the onsite basis $\left\{\ket{L},\ket{H_{+1}},\ket{H_{0}},\ket{H_{-1}}\right\}$ gives the following effective Hamiltonian,

\begin{align}
\mathcal{H}^{\mathrm{eff}}
=&
E_H\sum_{i}P_{Hi}
\notag
\\
&+
E_L\sum_{i}P_{Li}
\notag
\\
&+
\delta E_{LL}
\sum_{\langle ij \rangle}
P_{Li}P_{Lj}
\notag
\\
&+
\delta E_{HL}
\sum_{\langle ij \rangle}
\left(
P_{Hi}P_{Lj}+P_{Li}P_{Hj}
\right)
\notag
\\
&+
J_s
\sum_{\langle ij \rangle}
\left(
\vec{s}_i\cdot\vec{s}_j-1
\right)
P_{Hi}
P_{Hj}
\notag
\\
&+
J'
\sum_{\langle ij \rangle}
\left[
P^{-}_i
\left(
\vec{s}_i\cdot\vec{s}_j+1
\right)
P^{+}_j
+
\mathrm{H.c.}
\right]
\notag
\\
&+
I'
\sum_{\langle ij \rangle}
\left[
\left(
\vec{s}_i\cdot\vec{s}_j-1
\right)
P^{+}_i
P^{+}_j
+
\mathrm{H.c.}
\right].
\label{eq:h_eff_p}
\end{align}
The operators in Eq. (\ref{eq:h_eff_p}) are given as
\begin{align}
P_{Li}&=\ket{L}_i\bra{L}_i,
\\
P_{Hi}&=\sum_{s^z=-1,0,1}\ket{H_{s^z}}_i\bra{H_{s^z}}_i,
\\
P^{+}_i&=\ket{H_{0}}_i\bra{L}_i,
\\
P^{-}_i&=\ket{L}_i\bra{H_{0}}_i.
\end{align}
The factors in Eq. (\ref{eq:h_eff_p}) are given as
\begin{align}
E_{H}&=
\Delta+U'-J,
\\
E_{L}&=
\Delta+U
-\Delta',
\\
\delta E_{LL}&=
\frac{4f^2g^2(t_a^2+t_b^2)}{2U'-U-J+2\Delta'},
\\
\delta E_{HL}&=
(t_a^2+t_b^2)
\left[
\frac{f^2}{-\Delta+U'+\Delta'}
+
\frac{g^2}{\Delta+U'+\Delta'}
\right],
\\
J_s&=
\frac{t_a^2+t_b^2}{U+J},
\\
J'&=
2 t_a t_b
\left[
\frac{f^2}{-\Delta+U'+\Delta'}
+
\frac{g^2}{\Delta+U'+\Delta'}
\right],
\\
I'&=
2 t_a t_b f g
\left[
\frac{1}{U+J}
+
\frac{1}{2U'-U-J+2\Delta'}
\right].
\end{align}
We introduce a spin and pseudospin representation\cite{PhysRevB.93.205136} of Eq. (\ref{eq:h_eff_p}) to clarify the anisotropy of the Hamiltonian. First, we introduce descending and ascending operators between the $s=0$ LS state and $s=1$ HS quadrupoles, which are defined as
\begin{align}
P_{X}^{-}
&=
\frac{1}{\sqrt{2}}(
\ket{L}\bra{H_{-1}}
-
\ket{L}\bra{H_{+1}}
),
\label{eq:PXm}
\\
P_{Y}^{-}
&=
\frac{1}{\sqrt{2}i}(
\ket{L}\bra{H_{-1}}
+
\ket{L}\bra{H_{+1}}
),
\\
P_{Z}^{-}
&=\ket{L}\bra{H_{0}},
\\
P_{\Gamma}^{+}
&=
(P_{\Gamma}^{-})^{\dagger}
\quad
(\Gamma=X,Y,Z).
\label{eq:PGp}
\end{align}
Second, the pseudospin operators $\tau_{\Gamma}^{\gamma}$ for $\Gamma=X,Y,Z$ are defined with the descending and ascending operators in Eqs. (\ref{eq:PXm})-(\ref{eq:PGp}) as
\begin{align}
\tau_{\Gamma}^{x}
&=
P_{\Gamma}^{-}+P_{\Gamma }^{+},
\\
\tau_{\Gamma}^{y}
&=
i(P_{\Gamma}^{-}-P_{\Gamma }^{+}),
\\
\tau_{\Gamma}^{z}
&=
P_{\Gamma}^{+}P_{\Gamma}^{-}-P_L.
\end{align}
Note that the representations of the pseudospin operators $\tau_\Gamma^\gamma$ are twice as much as the ones of the spin 1/2 operator.
Finally, the spin and pseudospin representation of the effective Hamiltonian in Eq. (\ref{eq:h_eff_p}) is 
\begin{align}
\mathcal{H}^{\mathrm{eff}}
=&
E_{0\tau}
-h_{\tau z}
\sum_i
\tau^z_{i}
+J_s
\sum_{\langle ij \rangle}
\vec{s}_i \cdot \vec{s}_j
+
J_z
\sum_{\langle ij \rangle}
\tau^{z}_{i}
\tau^{z}_{j}
\notag \\
&+
J_x
\sum_{\Gamma=X,Y,Z}
\sum_{\langle ij \rangle}
\tau^{x}_{\Gamma i}
\tau^{x}_{\Gamma j}
+
J_y
\sum_{\Gamma=X,Y,Z}
\sum_{\langle ij \rangle}
\tau^{y}_{\Gamma i}
\tau^{y}_{\Gamma j},
\label{eq:h_eff_st}
\end{align}
where $\vec{s}$ is the spin operator whose amplitude is $s=1$, and $\tau_i^{z}=\tau_{Xi}^{z}+\tau_{Yi}^{z}+\tau_{Zi}^{z}$. Note that the spin operator $\vec{s}$ is generally not commutative with the pseudospin operator $\tau_{\Gamma}^{\gamma}$. The factors in Eq. (\ref{eq:h_eff_st}) are given as
\begin{align}
E_{0\tau}
&=
NE_{L}
-\frac{Nn_z\delta E_{LL}}{2}
+\frac{3N\tilde{\Delta}}{4}
+\frac{9Nn_zJ_n}{32},
\\
-h_{\tau z}&=
-\left(
\frac{\tilde{\Delta}}{4}
+
\frac{3n_zJ_n}{16}
\right),
\\
J_x
&=
\frac{J'+I'}{2},
\\
J_y
&=
\frac{J'-I'}{2},
\\
J_z
&=
\frac{J_n}{16},
\end{align}
where 
\begin{align}
\tilde{\Delta}
&=
E_H-E_L+n_z(\delta E_{LL}-\delta E_{HL}),
\\
J_n
&=
2\delta E_{HL}-\delta E_{LL} -J_s,
\end{align}
where $n_z$ is number of the nearest sites, and $z=2$ in the 1D system. 
When $I$ is nonzero in Eq. (\ref{eq:tohm}), then $J_x>J_y$ in Eq. (\ref{eq:h_eff_st}), and the number of excitons (HS states) does not conserve. Naively, the anisotropy of $J_x>J_y$ in the pseudospin-pseudospin interaction seems to induce a long-range order for $\braket{\tau_{\Gamma i}^x}$. However, since there are spin-spin and the other pseudospin-pseudospin interactions in Eq. (\ref{eq:h_eff_st}), the long-range order for $\braket{\tau_{\Gamma i}^x}$ can be suppressed as a result of the competition of the interactions. In fact, no symmetry breakings of $\braket{\tau_{\Gamma i}^x}$ are obtained in this study.
In addition to Eq. (\ref{eq:h_eff_st}), the Zeeman energy term is introduced as
\begin{align}
\mathcal{H}_{\mathrm{Zeeman}}=-H\sum_{i}s_i^{z}.
\end{align}
\par
Before showing the results of the numerical analysis in Sect. \ref{sec:results}, we show in Table \ref{ta:schematic_picture} the typical configurations of pseudospins in the effective model in Eq. (\ref{eq:h_eff_st}), where spin degrees of freedom and quantum fluctuations are not taken into consideration.
\begin{table}[htbp]
  \caption{Typical configurations of pseudospins in effective model in Eq. (\ref{eq:h_eff_st}).}
  \label{table:data_type}
  \centering
  \begin{tabular}{ccl}
    \hline
    Phase  & $\tau_{\Gamma}$ configurations  &  Parameter conditions  \\
    \hline \hline
    LS  &
    $\downarrow\;\;\;\downarrow\;\;\;\downarrow\;\;\;\downarrow$ &
    $h_{\tau z}\ll 0\;(\Leftrightarrow \Delta \gg  U-U'+J)$
    \\
    HS  &
    $\uparrow\;\;\;\uparrow\;\;\;\uparrow\;\;\;\uparrow$ &
    $h_{\tau z}\gg 0\;(\Leftrightarrow \Delta \ll  U-U'+J)$
    \\
    EI & 
    $\rightarrow\;\leftarrow\;\rightarrow\;\leftarrow$ &
    $h_{\tau z}\sim 0,\;J_x>J_y>J_z$
    \\
    & & $(\Leftrightarrow \Delta\sim U-U'+J,\; t_a/t_b\sim 1)$
    \\
    SSO &
    $\uparrow\;\;\;\downarrow\;\;\;\uparrow\;\;\;\downarrow$ &
    $h_{\tau z}\sim 0,\;J_x,J_y\ll J_z$
    \\
    & & $(\Leftrightarrow \Delta\sim U-U'+J,\; t_a/t_b\sim 0)$
    \\
    \hline
  \end{tabular}
  \label{ta:schematic_picture}
\end{table}
\par
In the case of $\Delta\gg U-U'+J$, where the crystal-field splitting is sufficiently large,
$h_{\tau z} \ll 0$ in the effective model in Eq. (\ref{eq:h_eff_st}), and the ground state is in a low-spin (LS) phase with 
($\tau_{\Gamma}:\; \downarrow\;\downarrow\;\dots\;\downarrow$).
In the case of $\Delta\ll U-U'+J$ and $h_{\tau z} \gg 0$, the ground state is in a high-spin (HS) phase with 
($\tau_{\Gamma}:\; \uparrow\;\uparrow\;\dots\;\uparrow$).
In the case where $\Delta$ and $U-U'+J$ are comparable, the inter-pseudospin interaction $J_\gamma$ and the inter-spin interaction $J_s$ are dominant. 
The inter-pseudospin interaction $J_\gamma$ is anisotropic for $\gamma=x,y,z$. Since $J_z\propto t_a^2+t_b^2$ and $J_x,J_y\propto 2t_a t_b$, the behavior of the ground state depends on the parameter $\eta=2t_a t_b/(t_a^2+t_b^2)$, which characterizes the anisotropy of the pseudospin-pseudospin interaction. The range of $\eta$ is $0\leq\eta\leq 1$, where $t_a/t_b=0$ for $\eta=0$, and $t_a/t_b=1$ for $\eta=1$. In the case of $\eta\ll 1$, the anisotropy becomes $J_z \gg J_x, J_y$, and the ground state tends to show a SSO with ($\tau_{\Gamma}:\; \uparrow\;\downarrow\;\dots\;\uparrow\;\downarrow$). 
If $\eta$ is not small, the anisotropy can be $J_z < J_y<J_x$, and pseudospins tend to form a state in which the pseudospin points in the $x$ axis, ($\tau_\Gamma:\;\rightarrow\;\leftarrow\;\dots\;\rightarrow\;\leftarrow$). In this configuration, the LS and HS states are hybridized.
Since the HS state is regarded as a spin-triplet exciton, the configuration ($\tau_\Gamma:\;\rightarrow\;\leftarrow\;\dots\;\rightarrow\;\leftarrow$) represents the EI configuration, in which the excitons move coherently.
In practice, the picture of the ground state of the effective model becomes more complicated due to the competition of multiple interactions in Eq. (\ref{eq:h_eff_st}), which include the spin-spin interaction and three types of pseudospin-pseudospin interactions identified by $\Gamma=X,Y,Z$.\par
We apply the DMRG method to the effective model. The system is 1D and has the open boundary condition unless otherwise specified. We set the truncation number in DMRG to $\chi = 256$ (and exceptionally $\chi=64$ to determine some phase boundaries). Under this setting, the truncation error of the reduced density matrix is less than $10^{-5}$. We additionally use the periodic boundary condition for analyses of the SSO phase, where incommensurate spin structures are found.

\section{Results} \label{sec:results}

\subsection{Ground-state phase diagram}

\begin{figure}[t]
\begin{center}
\includegraphics[width=7cm]{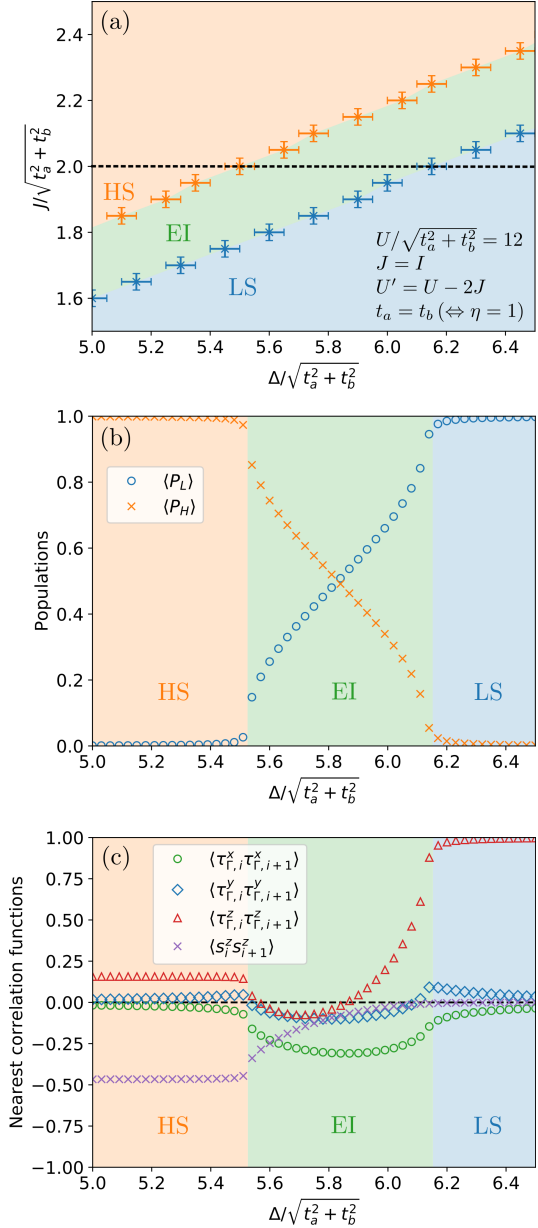}
\end{center}
\caption{
(Color online)
Panel (a) shows a ground-state phase diagram for the effective Hamiltonian in Eq. (\ref{eq:h_eff_st}) in the plane of the Hund coupling $J/\sqrt{t_a^2+t_b^2}$ and the crystal-field splitting $\Delta/\sqrt{t_a^2+t_b^2}$, where we set the parameters as $U/\sqrt{t_a^2+t_b^2}=12$ and $\eta=1\,(\Leftrightarrow t_a=t_b)$.
Panel (b) shows the populations of the LS and HS states along the dashed line in panel (a), where $J/\sqrt{t_a^2+t_b^2}=2$.
Panel (c) shows the nearest correlation functions along the dashed line in panel (a).
}
\label{fig:dj_phase_diagram}
\end{figure}
\begin{figure}[t]
\begin{center}
\includegraphics[width=7cm]{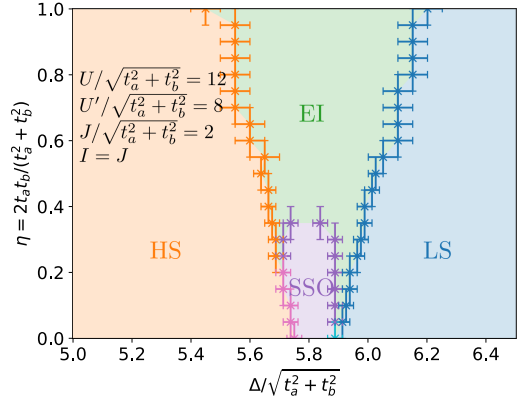}

\end{center}
\caption{
(Color online)
A ground-state phase diagram for the effective Hamiltonian in Eq. (\ref{eq:h_eff_st}) in the plane of the anisotropy factor for pseudospin $\eta$ and the crystal-field splitting $\Delta/\sqrt{t_a^2+t_b^2}$, where we fix that $U/\sqrt{t_a^2+t_b^2}=12$ and $J/\sqrt{t_a^2+t_b^2}=2$. The error bars represent discrete intervals in the analysis.
}
\label{fig:ej_phase_diagram}
\end{figure}

In this subsection, we present phase diagrams obtained by the numerical analysis with DMRG.\par
First, we fix the anisotropy factor for the pseudospin interaction at $\eta=1\,(\Leftrightarrow t_a=t_b)$ and make a phase diagram in the plane of the Hund coupling $J$ and the crystal-field splitting $\Delta$, where we set the parameters of the Coulomb interactions to $U/\sqrt{t_a^2+t_b^2}=12$, $U'=U-2J$, and $J=I$ as a strongly correlated region.
The $\Delta$-$J$ phase diagram is shown in Fig. \ref{fig:dj_phase_diagram}(a). 
In Fig. \ref{fig:dj_phase_diagram}(a), as $\Delta$ increases, the phase shifts as HS $\rightarrow$ EI $\rightarrow$ LS, where HS, EI, and LS stand for a low-spin phase, an excitonic insulating phase, and a high-spin phase, respectively.
We obtain the phase boundaries by the rapid changes of the entanglement entropy, which is discussed in SubSect. \ref{sec:results}.4. 
\par

In Fig. \ref{fig:dj_phase_diagram}(b), we present the populations of the HS state $\braket{P_H}$ and LS state $\braket{P_L}$ along the dashed line in Fig. \ref{fig:dj_phase_diagram}(a). In the EI phase, the LS state and the HS state coexist, and the population of the LS state increases as $\Delta$ increases. 
\par

We can understand the property of the EI phase in the nearest correlation functions. We show the nearest correlation functions,  $\braket{\tau_{\Gamma i}^x \tau_{\Gamma i+1}^x}$,
$\braket{\tau_{\Gamma, i}^y \tau_{\Gamma, i+1}^y}$,
$\braket{\tau_{\Gamma, i}^z \tau_{\Gamma, i+1}^z}$, and
$\braket{s_{i}^z s_{i+1}^z}$
in Fig. \ref{fig:dj_phase_diagram}(c). Note that the product of the nearest pseudospin, $\braket{\tau_{\Gamma i}^z}\braket{\tau_{\Gamma i+1}^z}$, is not subtracted from $\braket{\tau_{\Gamma i}^z \tau_{\Gamma i+1}^z}$ although $\braket{\tau_{\Gamma j}^z}$ is positive in the HS phase and negative in the LS phase. 
We compute the correlation function at the center of the system.
The values of the pseudospin correlations are independent of the quadrupole basis $\Gamma=X, Y, Z$ because of the rotational symmetry. 
The large value of $|\braket{\tau_{\Gamma,i}^x \tau_{\Gamma, i+1}^x}|$ in the EI phase indicates that the EI phase is stabilized by the local kinetic energy of excitons.
Compared with $\braket{\tau_{\Gamma,i}^{x} \tau_{\Gamma,i+1}^{x}}$, $\braket{\tau_{\Gamma,i}^{y} \tau_{\Gamma,i+1}^{y}}$ does not show strong antiferrmagnetism.
This is because the antiferromagnetic interactions between pseudospins satisfy $J_x>J_y$.
$\braket{\tau_{\Gamma,i}^{z} \tau_{\Gamma,i+1}^{z}}$ also does not show strong antiferromagnetism even near the center of the EI phase as a result of $J_x>J_z$. 
Near the HS(LS) phase, large value of $-h_{\tau z}$ $(h_{\tau z})$ makes $\braket{\tau_{\Gamma,i}^{z} \tau_{\Gamma,i+1}^{z}}$ ferromagnetic.
In terms of long-range correlation functions, 
$\braket{\tau_{\Gamma i}^{x}\tau_{\Gamma j}^{x}}$
and
$\braket{\tau_{\Gamma i}^{y}\tau_{\Gamma j}^{y}}$
become zero at $|i-j|\rightarrow \infty$. 
This means that there is no long-range order in the EI phase, that is $\braket{\tau_{\Gamma i}^{x}}=\braket{\tau_{\Gamma i}^{y}}=0$ for $\Gamma=X,Y,Z$ and $i$ at any site.
Since there are spin-spin interaction and three types of pseudospin-pseudospin interactions identified by $\Gamma$ in Eq. (\ref{eq:h_eff_st}),
the long-range order of EI can be suppressed as a result of the competition of the interactions.

The nearest spin correlation $\braket{s_{i}^z s_{i+1}^z}$ in Fig. \ref{fig:dj_phase_diagram}(c) is antiferromagnetic in the HS and EI phases.
In the HS phase, $\braket{P_H}\sim 1$. In this region, the effective Hamiltonian reduces to the antiferromagnetic spin-1 Heisenberg model, and the nearest spin correlations becomes negative.
In the EI phase, $\braket{P_H}$ is finite, and antiferromagnetic spin-spin interaction between the HS states make the nearest spin correlation negative.
\par
Next, we present a phase diagram in the plane of the anisotropy for the pseudospin interaction $\eta$ and the crystal-field splitting $\Delta$ in Fig. \ref{fig:ej_phase_diagram}, where we fix the parameters of Coulomb interactions to $U/\sqrt{t_a^2+t_b^2}=12$, $U'/\sqrt{t_a^2+t_b^2}=8$, and $J/\sqrt{t_a^2+t_b^2}=I/\sqrt{t_a^2+t_b^2}=2$.
In addition to the HS, EI, and LS phases, the SSO phase is found in Fig. \ref{fig:ej_phase_diagram}.
In the SSO phase, the spin states show superlattice structures, in which the population of HS states varies from site to site. The spatial structure of SSO changes depending on $\Delta$ and $\eta$. We focus on the spatial structures in the next subsection.
\par
We characterize the SSO phase on the basis of the existence of a superstructure of spin states.
The criterion is that a state is in the SSO phase when
$
\mathrm{max}(\braket{P_{Hi}})
-
\mathrm{min}(\braket{P_{Hi}})
>10^{-2}
$ for $i$ in the central 20 sites in a 60-sites system with open boundaries.\par
The outlines in the phase diagrams in Fig. \ref{fig:dj_phase_diagram} and Fig. \ref{fig:ej_phase_diagram} can be understood as the typical configurations of pseudospins which is shown in Table \ref{ta:schematic_picture}, and is consistent with the previous studies with mean-field approximation\cite{PhysRevB.93.205136} and dynamical mean-field approximation.
\cite{srep30510}
However, no spin or pseudospin orderings are observed in the HS and EI phases. This is interpreted as the result of the effect of strong quantum fluctuations in a 1D system, which is considered in the DMRG method.
In addition, various types of spin-state structures were obtained by using large 1D clusters.

\subsection{Spatial structures of spin states in the EI and SSO phases}

In this subsection, we present the spatial structures of the spin states in the EI phase and the SSO phase through correlation functions. We set the system size as $L=60$ and use the periodic boundary condition.\par
We consider a Fourier transformed correlation functions for spin states as
\begin{align}
C_{X}(k)
&=
\frac{1}{L}\sum_{ij}
\braket{X_i X_j}
\cos\left[k(i-j)\right],
\end{align}
where $X_i$ is a one-site operator at a site $i$. Note that the product of one-site expected values are not subtracted in $C_{X}(k)$.\par

\begin{figure}[t]
\begin{center}
\includegraphics[width=6.7cm]{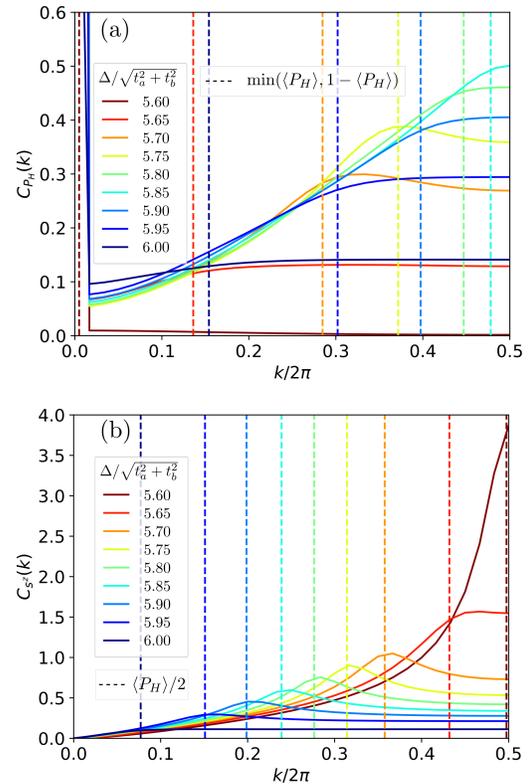}
\end{center}
\caption{
(Color online)
Fourier transformed correlation functions in the EI phase.
Panel (a) shows the correlations of HS state $C_{P_H}(k)$ at $\eta=0.5$ for some $\Delta/\sqrt{t_a^2+t_b^2}$. The dashed lines in panel (a) are the expected values of $\mathrm{min}(\braket{P_H},1-\braket{P_H})$ per site.
Panel (b) shows the spin correlation $C_{s^z}(k)$ at $\eta=0.5$. The dashed lines in panel (b) are $\braket{P_H}/2$.
}
\label{fig:ei_correlation_k}
\end{figure}
\begin{figure}[htbp]
\begin{center}
\includegraphics[width=6.7cm]{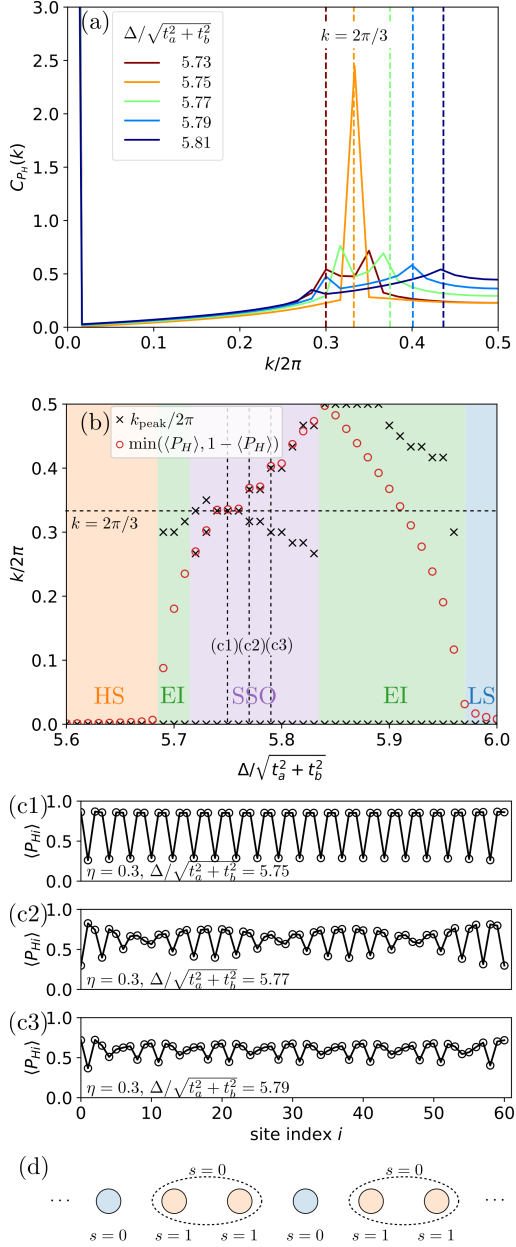}

\end{center}
\caption{
(Color online)
Fourier transformed correlation functions in the SSO phase.
Panel (a) shows the correlations of HS state $C_{P_H}(k)$ at $\eta=0.3$ for some $\Delta/\sqrt{t_a^2+t_b^2}$. The dashed lines are the expected values of $\mathrm{min}(\braket{P_H},1-\braket{P_H})$.
Panel (b) shows the peak wavenumber against the crystal-field splitting $\Delta/\sqrt{t_a^2+t_b^2}$.
In Panels (c1), (c2), and (c3), some SSO structures obtained in the SSO phase are shown. Each parameter corresponds to the dashed line in panel (b). 
Panel (d) shows the schematic picture of the LS/HS/HS structure in Panel (c1).
}
\label{fig:sso_correlation_k}
\end{figure}
First, we fix the parameters as $U/\sqrt{t_a^2+t_b^2}=12,J/\sqrt{t_a^2+t_b^2}=I/\sqrt{t_a^2+t_b^2}=2,U'=U-2J,\eta=0.5$. 
In this region, the phase transition is HS $\rightarrow$ EI $\rightarrow$ LS as $\Delta/\sqrt{t_a^2+t_b^2}$ increases.
\par
In Fig. \ref{fig:ei_correlation_k}(a), we present the correlation function for the HS state $C_{P_H}(k)$ in the EI phase.
Since $\braket{P_{Hi}}$ is uniform in the EI phase, $k$-dependence of $C_{P_H}(k>0)$ corresponds to the spatial quantum fluctuation in the population of HS states. $C_{P_H}(k=0)$ originates in the uniform populations of the HS state. 
$k$-dependence of $C_{P_H}(k>0)$ is small in the HS phase and the LS phase and becomes large in the EI phase.
This behavior means that, in the EI phase, there is a strong quantum fluctuation, and the ground state is far from the direct product state such as $(\tau_\Gamma: \rightarrow\;\leftarrow\dots\rightarrow\;\leftarrow)$, in which $C_{P_H}(k>0)$ is constant.
In Fig. \ref{fig:ei_correlation_k}(a),
we also show the wave number at $2\tilde{k}=\mathrm{min}(2\pi\braket{P_H},2\pi (1-\braket{P_H}))$ by broken lines, where $\braket{P_H}$ is the spatial average of the populations of HS states.
If excitons are regarded as free fermions, $\tilde{k}$ is interpreted as the Fermi wavenumber.
In the EI phase, $C_{P_H}(k>0)$ tends to rise as $k$ increases in $k<\tilde{k}$ and becomes constant in $k>\tilde{k}$.
This behavior is similar to the behavior of 1D spinless free fermions. As a difference
from the case of fermions, $C_{P_H}(k \rightarrow 0)$ is non-zero in the EI phase. This originates in the anisotropy of $J_x>J_y$ in Eq. (\ref{eq:h_eff_st}), which means the number of excitons does not conserve.

\par
In Fig. \ref{fig:ei_correlation_k}(b), we show the Fourier transformed spin correlation $C_{s^z}(k)$ with the same parameters as those in Fig. \ref{fig:ei_correlation_k}(a).
$C_{s^z}(k)$ has a peak at $k=\pi\braket{P_H}$ in the EI and HS phases.
This is interpreted as the spin correlation in the EI phase representing a spin density wave whose wavenumber corresponds to the population of the HS state, although there is no spin ordering.
\par

Next, we fix $\eta=0.3$. In this region, the phase transitions are HS $\rightarrow$ EI $\rightarrow$ SSO $\rightarrow$ EI $\rightarrow$ LS as $\Delta/\sqrt{t_a^2+t_b^2}$ increases. 
Figure \ref{fig:sso_correlation_k}(a) shows $C_{P_H}(k)$ in the SSO phase. We also show the peak wavenumbers against the crystal-field splitting in Fig. \ref{fig:sso_correlation_k}(b).
The sharp peaks in Fig. \ref{fig:sso_correlation_k}(a) represent the structure of the SSO. 
There is a strong peak for $k=2\pi/3$ in $\Delta/\sqrt{t_a^2+t_b^2}=5.75$. This peak represents the LS/HS/HS structure shown in Fig. \ref{fig:sso_correlation_k}(d).
\par
As shown in Fig. \ref{fig:sso_correlation_k}(b), the spin-state structure in the SSO phase changes against the crystal-field splitting $\Delta/\sqrt{t_a^2+t_b^2}$.
The plateau at $k=2\pi/3$ in Fig. \ref{fig:sso_correlation_k}(b) indicates that the LS/HS/HS structure is stable. According to the Hamiltonian in Eq. (\ref{eq:h_eff_st}), the nearest LS-HS pair is stabilized by the $J_z$ term, and the nearest HS-HS pair is stabilized by the $J_s$ term. The LS/HS/HS  structure is realized as a result of the competition between these interactions.
In the other region in the SSO phase, two peaks branch from the peak at $k = 2\pi/3$. 
The two peaks have generally incommensurate wavenumbers, and one of them follows the population of the HS state $\braket{P_H}$. The discretization of the $k_{\mathrm{peak}}$ in Fig. \ref{fig:sso_correlation_k}(b) seems to originate from the finite system size $L=60$ and should become continuous in $L\rightarrow \infty$.
Some snapshots of the spin-state structures in the SSO phase are shown in Fig. \ref{fig:sso_correlation_k}(c1)-(c3). 
Figure \ref{fig:sso_correlation_k}(c1) shows the LS/HS/HS structure shown in Fig. \ref{fig:sso_correlation_k}(d). 
Figure \ref{fig:sso_correlation_k}(c2) shows the structure at $\Delta/\sqrt{t_a^2+t_b^2}=5.77$, which is near the parameter at which LS/HS/HS structure realizes. Here some kinks are inserted into the basic LS/HS/HS structure. 
Figure \ref{fig:sso_correlation_k}(c3) shows the structure at $\Delta/\sqrt{t_a^2+t_b^2}=5.79$, which is far from the parameter of LS/HS/HS structure. Here a complex spin state is found.
In the region where $\braket{P_H}<0.5$, the SSO phase is not found. This result indicates that the spin-spin interaction between the HS states contributes to the stabilization of the SSO with the LS/HS/HS structure.

\subsection{Magnetic-field effects}
\begin{figure*}[t]
\begin{center}
\includegraphics[width=14cm]{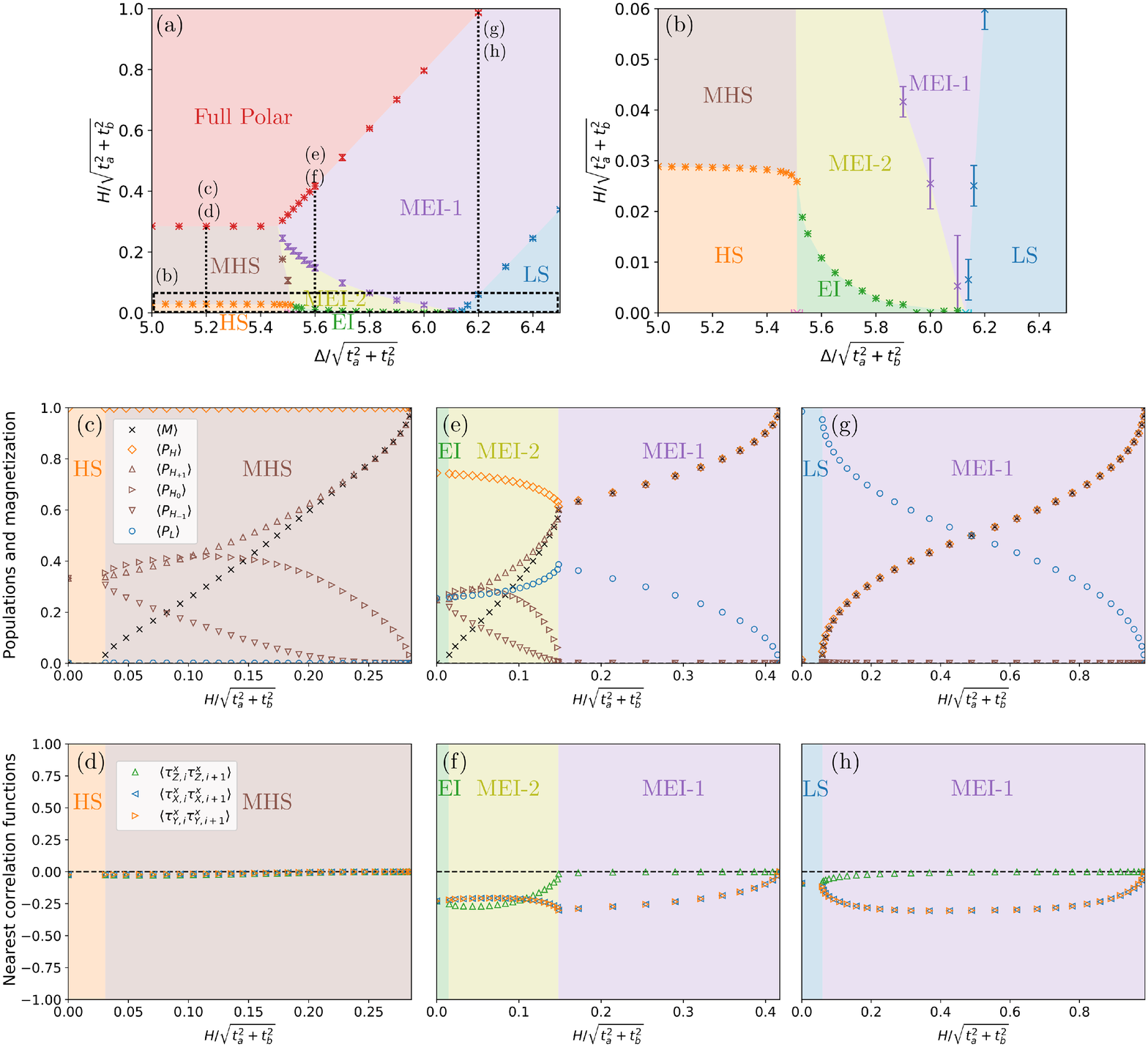}
\end{center}
\caption{
(Color online)
Phase diagram of the effective model in a magnetic field. Panel (a) is phase diagram in the plane of a magnetic field  $H/\sqrt{t_a^2+t_b^2}$ and crystal-field splitting $\Delta/\sqrt{t_a^2+t_b^2}$. 
Panel (b) is the enlarged view of the low magnetic-field region enclosed by the dashed line in panel (a).
The error bars in Panels (a) and (b) represent the discretization of the magnetization in the finite system size $L=60$. Exceptionally, the phase boundaries at the HS-MHS and the EI-MEI-2 are determined by the extrapolation. 
The dashed lines in panel (a) represent the magnetization processes in panels (c)-(h). Panels (c), (e), and (g) shows the magnetization curve and the populations of spin states in the magnetization process in the HS, EI, and LS phases, respectively. Panels (d), (f), and (h) show nearest pseudospin correlations in the magnetization process in the HS, EI, and LS phases, respectively.
}
\label{fig:magnetic_field_effects}
\end{figure*}

\begin{figure}[htbp]
\begin{center}
\includegraphics[width=7cm]{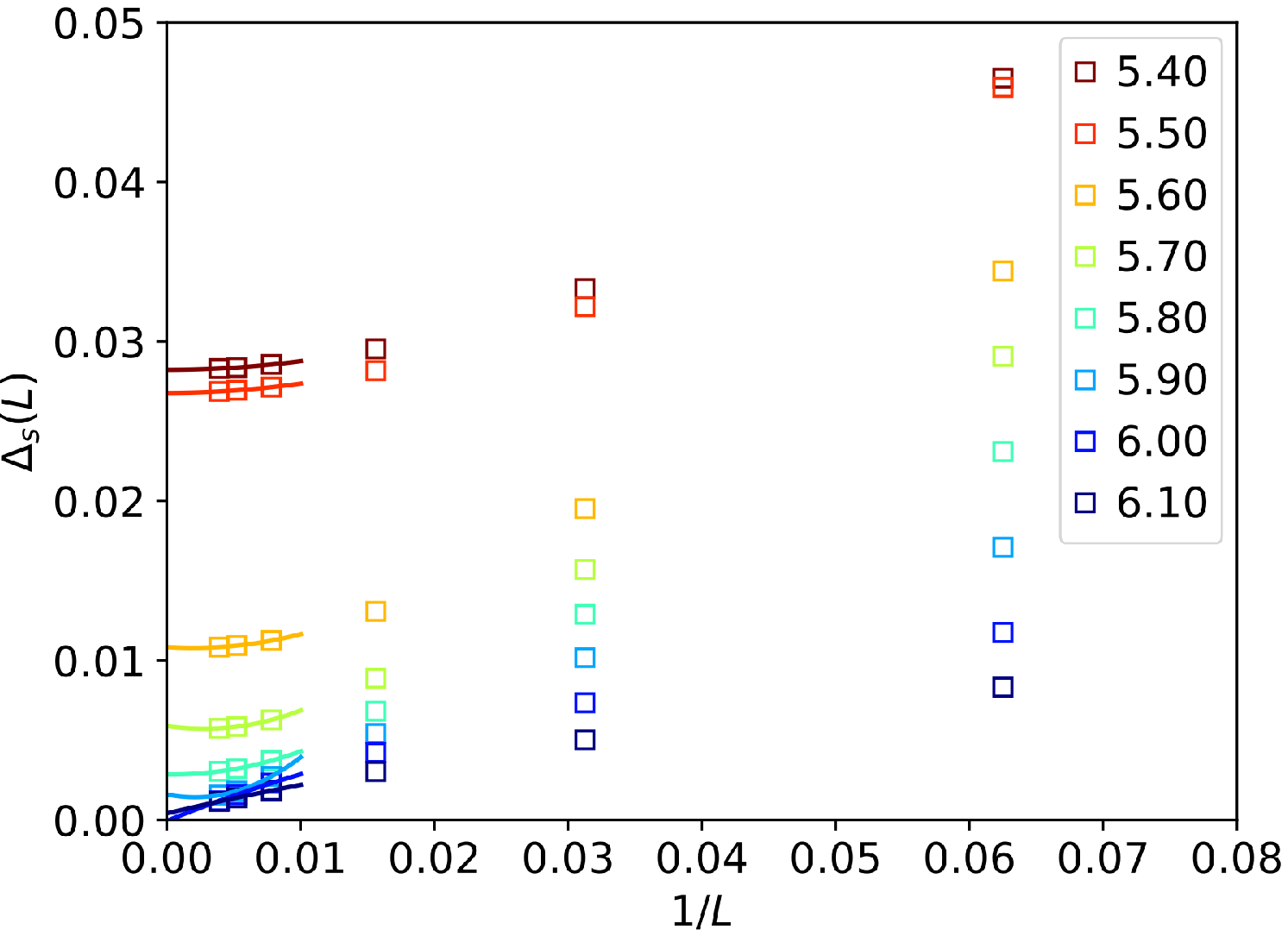}
\end{center}
\caption{
(Color online)
Samples of Spin gap extrapolations with Eq. (\ref{eq:delta_s_ex}).
$\Delta/\sqrt{t_a^2+t_b^2}=5.40,5.50$ are in the HS phase, and the others are in the EI phase. $\Delta/\sqrt{t_a^2+t_b^2}=5.90,6.00,6.10$ are near the LS phase.
}
\label{fig:spin_gap_extrapolation}
\end{figure}

In this subsection, we present the magnetic-field effects in the effective model. Figures \ref{fig:magnetic_field_effects}(a) and \ref{fig:magnetic_field_effects}(b) are phase diagrams in the plane of the crystal-field splitting $\Delta/\sqrt{t_a^2+t_b^2}$ and a magnetic field $H/\sqrt{t_a^2+t_b^2}$, where we fix the parameters, $U/\sqrt{t_a^2+t_b^2}=12$, $J/\sqrt{t_a^2+t_b^2}=I/\sqrt{t_a^2+t_b^2}=2$, $U'=U-2J$ and $\eta=1\,(\Leftrightarrow t_a=t_b)$. In these parameters, the EI phase exists in the intermediate region between the HS and LS phases.
In Figs.\ref{fig:magnetic_field_effects}(a) and \ref{fig:magnetic_field_effects}(b), in addition to the 
HS, EI, and LS phases, there appears a type-1 magnetized excitonic insulating phase (MEI-1), a type-2 magnetized excitonic insulating phase (MEI-2), a magnetized high-spin phase (MHS), and a fully polarized phase (Full Polar).
We determine the phase boundary on the basis of the discontinuity of the first derivative of the magnetization curve.
Basically, we set the system size as $L=60$ and set the truncation number in DMRG as $\chi=64$. 
\par
The value of a magnetic field at the phase boundaries for HS-MHS, EI-MEI-2, and LS-MEI-1 corresponds to the spin gaps in HS, EI, and LS phases, respectively. As shown in Figs.\ref{fig:magnetic_field_effects}(a) and \ref{fig:magnetic_field_effects}(b), spin gaps are obtained in the HS, EI, and LS phases,
and the spin gaps appears to be closed at the EI-LS phase boundary.
To determine the spin gaps in the HS and EI phases, system-size extrapolations are performed. The extrapolation is performed with the function
\begin{align}
\Delta_s(L)=\Delta_s(\infty)+a/L+b/L^2,
\label{eq:delta_s_ex}
\end{align}
where $\Delta_s(\infty)$, $a$, and $b$ are the fitting parameters. $\Delta_s(L)$ is a spin gap in the system whose size is $L$. $\Delta_s(L)$ is computed by  $\Delta_s(L)=E^{(m=2)}(L)-E^{(m=0)}(L)$, where $E^{(m)}(L)$ is the energy of the ground state at $\braket{\sum_{i}s^z_i}=m$. We avoid $E^{(m=1)}(L)-E^{(m=0)}(L)$ because it corresponds to the spin excitation at the edges,
and this magnetization is ignored in an infinite system.
We use $L=128,192,256$ under the open boundary condition and set the truncation number in DMRG to $\chi=256$. Some examples for the $L$ dependence of the $\Delta_s(L)$ are shown in Fig. \ref{fig:spin_gap_extrapolation}. Figure \ref{fig:spin_gap_extrapolation}(a) and (b) show that there is a finite spin gap in the HS and EI phases.
In the EI phase near the LS phase of $5.90<\Delta/\sqrt{t_a^2+t_b^2}<6.15$, the spin gap becomes small and its existence is not obtained with sufficient accuracy. Therefore, we omit the points in this region.
\par

The spin gap in the HS phase is the Haldane-gap because in $\Delta \ll U-U'+J$, the low-energy effective Hamiltonian in Eq. (\ref{eq:h_eff_st}) reduces to the spin-1 Heisenberg model.
In fact, the value of the spin gap in the HS phase in Fig. \ref{fig:magnetic_field_effects}(b) is in good agreement with $\Delta_{\mathrm{Haldane}}J_s= 0.0293$, where $J_s=(t_a^2+t_b^2)/(U+J)=0.0714$ is the amplitude of spin-spin interaction in Eq. (\ref{eq:h_eff_st}),
and $\Delta_{\mathrm{Haldane}}=0.4105$ is the value of Haldane gap of the spin-1 Heisenberg model obtained in the previous study.\cite{PhysRevB.48.10345}
\par
The spin gap in the EI phase is similar to the one of the spin-1 Heisenberg model as follows.
First, the value of spin gap in the EI phase connects to the one in the HS phase.
Second, the ground state energies in the EI phase degenerate for $m=0,\pm 1$ under the open boundary condition, and the degeneracy lifts under the periodic boundary condition.
This result suggests the existence of free 1/2 spins at edges similar to the case of the AKLT model. We discuss the edge state in detail in Subsect. 3.5 through the observation of entanglement spectra.
\par
The spin gap in the LS phase is due to the energy difference between the LS state and the HS state.
\par
From here, we present the magnetization processes in the phases along the dashed lines (c)-(h) in Fig. \ref{fig:magnetic_field_effects}(a).
\par
The magnetization process in a wide range of the HS phase is HS $\rightarrow$ MHS $\rightarrow$ Full Polar (the dashed line at (c) and (d) in Fig. \ref{fig:magnetic_field_effects}(a)). 
In the MHS phase, the population of the LS state is almost zero, and $\braket{\tau_{\Gamma,i}^x\tau_{\Gamma,i+1}^x}$ is almost zero. 
Therefore, the magnetization process in the MHS phase is the process in which $s=1$ spins are aligned by a magnetic field.
\par
The magnetization process in the EI phase is EI $\rightarrow$ MEI-2 $\rightarrow$ MEI-1 $\rightarrow$ Full Polar (the dashed line at (e) and (f) in Fig. \ref{fig:magnetic_field_effects}(a)), where we call the MEI phase in a low magnetic field MEI-2 and the phase in a high magnetic field MEI-1.
As shown in Fig. \ref{fig:magnetic_field_effects}(e), in the MEI-2 phase, the population of the HS state is finite for all bases of $s^z=+1,0,-1$,
whereas in the MEI-1 phase, the populations of $s^z=+1$ is finite, and that of $s^z=0,-1$ is almost zero. 
Therefore, the phase transition of MEI-2 $\rightarrow$ MEI-1 occurs when the $s = 1$ spins are fixed to $s^z=+1$ by a magnetic field. Because of the antiferromagnetic spin-spin interaction, 
MEI-2 phase survives even in a finite magnetic field.
Regarding the nearest correlation functions shown in Fig. \ref{fig:magnetic_field_effects}(f), while $\braket{\tau_{\Gamma,i}^x \tau_{\Gamma,i+1}^x}$ for $\Gamma=X,Y,Z$ are all negative in the MEI-2 phase,  $\braket{\tau_{Z,i}^x \tau_{Z,i+1}^x}$ is almost zero in the MEI-1 phase. This means that the MEI-1 phase is stabilized by the kinetic energy of $s^z=+1$ excitons.
Since   
$\ket{H_{s^z=+1}}=-(P_{X}^{+}\ket{L}+iP_{Y}^{+}\ket{L})/\sqrt{2}$,
the $X$ and $Y$ excitons coexist with the $z$-direction magnetization, while the $Z$ excitons compete with the magnetization. 
As a result, in the MEI-2 phase, the $X$, $Y$, and $Z$ excitons are stabilized, while in the MEI-1 phase, only $s^z=+1$ exciton is stabilized.
\par
The magnetization process in the LS phase is LS $\rightarrow$ MEI-1 $\rightarrow$ Full Polar (the dashed line at (g) and (h) in Fig. \ref{fig:magnetic_field_effects}(a)). The MEI-1 phase is the same phase as in the magnetization process in the EI phase. As shown in Figs. \ref{fig:magnetic_field_effects}(g) and \ref{fig:magnetic_field_effects}(h), the properties of the populations and the nearest correlations are consistent with the MEI-1 phase in the magnetization process in the EI phase.

\par
In Fig. \ref{fig:magnetic_field_effects}(a),  we presented the magnetic-field effects around the EI phase, where $\eta=1\,(\Leftrightarrow t_a=t_b)$. 
The magnetic field effects for smaller $\eta$ are left for  future work, where the magnetized SSO phase may be found because the LS/HS phase was obtained under a magnetic field with a mean-field analysis.\cite{JPSJ.85.083706}

\subsection{Entanglement entropy}

\begin{figure}[htbp]
\begin{center}
\includegraphics[width=7cm]{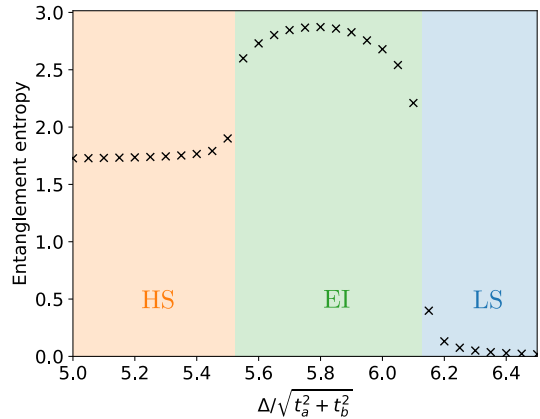}
\end{center}
\caption{
(Color online)
Entanglement entropy in the HS, EI, and LS phases, where the system whose size is $L=60$ is devided into the left and right one whose size is $l=30$.
The base of logarithm is $e$.
The model parameters are $\eta=1.0\,(\Leftrightarrow t_a=t_b)$, $U/\sqrt{t_a^2+t_b^2}=12$, $U'=U-2J$ and $J=I$.
}
\label{fig:entanglement_entropy_2}
\end{figure}
In this subsection, we present the properties of entanglement entropy in the HS, EI, and LS phases. Generally, entanglement entropy is sensitive to phase transitions, and is useful to determine the phase boundaries
\cite{PhysRevLett.112.026401,PhysRevB.92.075423,j.physleta.2015.09.034}
\par
Figure \ref{fig:entanglement_entropy_2} shows the entanglement entropy as a function of crystal-field splitting.
The entanglement entropy is defined for the density matrix of a $30$ site subsystem in a $60$ site system.
The system has the periodic boundary condition.
The truncation number in DMRG is set to $\chi=256$ for Fig. \ref{fig:entanglement_entropy_2}.
We set $\eta=1.0\,(\Leftrightarrow t_a=t_b)$, $U/\sqrt{t_a^2+t_b^2}=12$, $U'=U-2J$ and $J=I$, which are the same as the parameters in the dashed line in Fig. \ref{fig:dj_phase_diagram}(a).
\par
The value of entanglement entropy in the HS phase is consistent with the one of the spin-1 Heisenberg model.\cite{JPSJ.76.074603}
The entanglement entropy in the HS phase is not far from $2\log 2 \sim 1.39$. 
The value of $2\log 2$ is derived from the valencebond in the AKLT model.
The picture of valence bond solid (VBS) in the AKLT model\cite{PhysRevLett.59.799} is applied to the picture in the HS phase.
In this picture, the value of $2\log 2$ originates in the two valence bonds in the edges of the subsystem, and the additional value comes from the modulation from the AKLT model.
In the EI phase, the entropy is larger than the one in the HS and LS phases.  
The rapid changes in the entanglement entropy are found in the HS-EI and EI-LS boundaries.
Based on these changes, we determined the phase boundaries in Fig. \ref{fig:dj_phase_diagram} and Fig. \ref{fig:ej_phase_diagram}. 
In the LS phase, the entropy is quite small. This is because the electrons in the LS phase cannot move due to the Pauli exclusion principle.

\subsection{Entanglement spectra}
\begin{figure*}[t]
\begin{center}
\includegraphics[width=16cm]{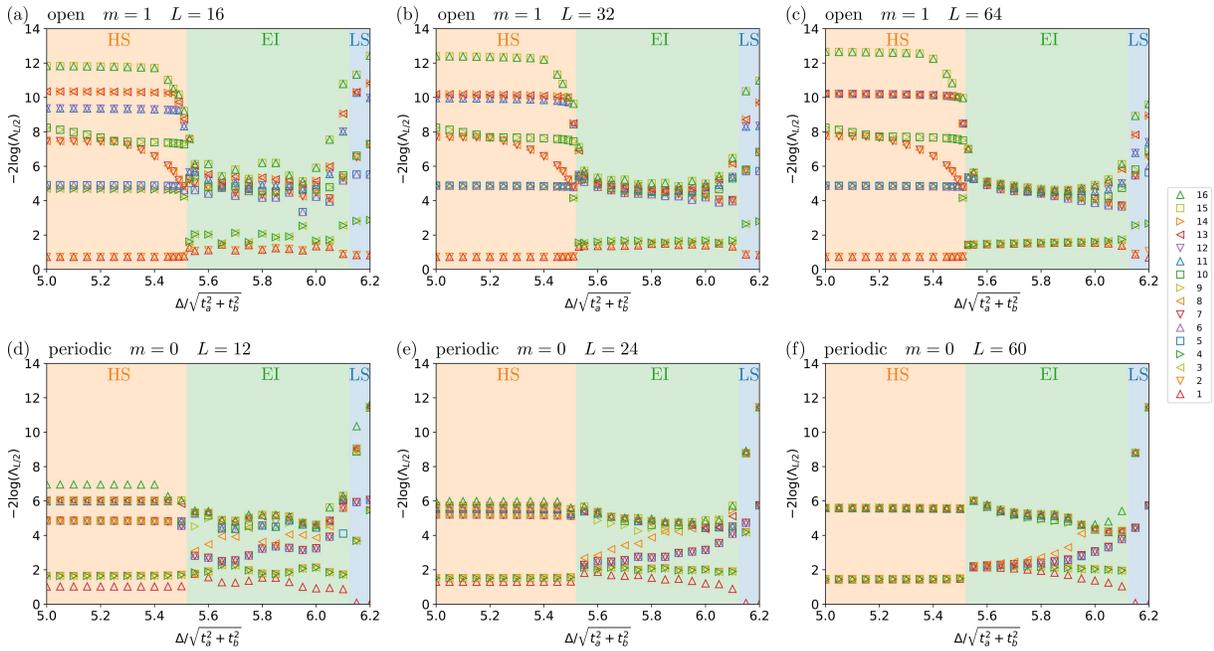}
\end{center}
\caption{
(Color online)
Entanglement spectra. 
The vertical axis $-\log(\Lambda_{L/2})$ means the entanglement spectra, where the divided systems have the same number of sites.
The base of the logarithm is Napier's constant.
For each parameter of $\Delta/\sqrt{t_a^2+t_b^2}$, we show 16 entanglement spectra whose singular values are the largest ones.
At panels (a), (b), and (c), the boundary conditions are open ones, and $m=\langle \sum_{i}s_i^z\rangle=1$.
The system sizes are (a) $L=16$, (b) $L=32$, and (c) $L=64$. 
At panels (d), (e), and (f), the boundary conditions are periodic ones, and $\langle \sum_{i}s_i^z\rangle=0$.
The system sizes are (d) $L=12$, (e) $L=24$, and (f) $L=60$. 
}
\label{fig:entanglement_spectra}
\end{figure*}

\begin{figure}[htbp]
\begin{center}
\includegraphics[width=7cm]{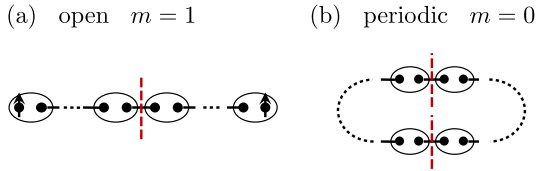}
\end{center}
\caption{
(Color online)
Schematic VBS pictures in the AKLT model.
Panel (a) shows the picture for $m=1$ under the open boundary condition, where $m=\braket{\sum_i s_i^z}$.
Panel (b) shows the picture for $m=0$ under the periodic boundary condition. 
The hollow circles represent sites that have $s=1$ spins.
The filled circles represent 1/2 spins, which form a singlets represented by the solid lines.
The arrows at edges represent 1/2 spins.
The red dashed lines divide the system into two sides.
}
\label{fig:vbs_hsei}
\end{figure}

In this section, we discuss the quantum state in the EI phase through the observation of entanglement spectra. 
As described in SubSect. 3.3, spin gaps are obtained in the HS and EI phases, which suggests the relationship with the Haldane gap.
\par
In the case of the spin-1 Heisenberg model, it is known that the entanglement spectra with the largest singular values are degenerate.\cite{PhysRevB.81.064439}
This degeneracy comes from the fact that the ground state is understood by the VBS picture in the AKLT model. In the VBS picture, singlet composed by virtual 1/2 spins results in the degeneracy of $\uparrow\downarrow$ and $\downarrow\uparrow$.
\par
In Fig. \ref{fig:entanglement_spectra}, we show the entanglement spectra for the HS, EI, and LS phases in our model. The parameters are set to $U/\sqrt{t_a^2+t_b^2}=12$,
$J/\sqrt{t_a^2+t_b^2}=2$,
$I/\sqrt{t_a^2+t_b^2}=2$,
$U'=U-2J$
and
$t_a=t_b$, which are the parameters at the $H=0$ line in Fig. \ref{fig:magnetic_field_effects}(a).
As $\left(\mathrm{boundary}\;\mathrm{condition}, m=\braket{\sum_i s_i^z}\right)$, we show the cases of (open, $m=1$) and (periodic, $m=0$). The reasons for adopting these settings are described later in this subsection. For the system sizes, we adopt $L=16,32,64$ for the open boundary condition, and $L=12,24,60$ for the periodic one.
\par
In the HS phase, the entanglement spectra for the maximum singular values degenerate to 2-fold under the open boundary condition as the system size increases as (a)$\rightarrow$(b)$\rightarrow$(c), and degenerate to 4-fold under the periodic boundary condition as (d)$\rightarrow$(e)$\rightarrow$(f).
These degeneracies are explained by VBS pictures in Figs. \ref{fig:vbs_hsei}(a) and \ref{fig:vbs_hsei}(b).
When the system is divided, the degeneracy in the entanglement spectra appears due to the singlet composed of 1/2 spins on the cross sections of both sides.
Under the open boundary condition, one set of singlets is divided, and under the periodic one, two sets of singlets are divided.
Therefore, there appears 2-folded degeneracy under the open boundary condition and  4-folded degeneracy under the periodic boundary condition.
The reason why we adopt the condition of $m = 1$ under the open boundary condition is to align the almost free 1/2 spins at both edges in the same direction.
These results for the entanglement spectra in the HS phase are consistent with previous studies in the spin-1 Heisenberg model.\cite{PhysRevB.91.045121,PhysRevB.81.064439}
\par
In the EI phase, as the system size increases, the spectra for the maximum singular values degenerate to 4-fold under the open boundary condition and degenerate to 8-fold under the periodic boundary condition. 
These degeneracies are explained as follows.
First, as in the case of the HS phase, singlets at edges form entanglement of 2-folded degeneracy in the open boundary condition and 4-folded degeneracy in the periodic one. 
Second, in the EI phase, there are contributions to entanglements due to motions of excitons and fluctuations in the number of excitons. 
In our model in Eq. (\ref{eq:h_eff_p}), we can obtain the following commutation relation $[P_H^{(2)},\mathcal{H}^{\mathrm{eff}}]=0$ where 
$P_H^{(2)}$ is a projection operator to the space where the total number of excitons is odd.
$P_H^{(2)}$ is represented by
$P_H^{(2)}=\sum_{\Lambda} \left(\prod_{\lambda\in\Lambda} P_{\lambda}^{H}\right)$, where $\Lambda$ denotes any set of site indexes with an odd number of elements. 
We introduce $N_H^{(2)}=0,1$ as a engenvalues of  $P_H^{(2)}$. 
The eigenstate of $N_H^{(2)}=0$ indicates the superposition of the states where the number of excitons is even, and the other indicates the odd one.
We consider the case when whole system is divided into two parts where $N_H^{(2)}$ of the whole system is fixed to 0. 
This condition is used in obtaining Fig. \ref{fig:entanglement_spectra}.
In this case, the possible combinations of $N_H^{(2)}$ of subsystems are $(N_H^{(2)L},N_H^{(2)R})=(0,0)$ or $(1,1)$.
If the sizes of the both subsystems and the number of excitons are large enough, the two entanglement spectra originating from $(0,0)$ and $(1,1)$ are degenerate.
The degree of degeneracy of this entanglement spectra is not affected by boundary conditions since $N_H^{(2)}$ is a property of bulk.

From the above, the combination of singlets in the valence bonds and fluctuations of the excitons explains
the 4(8)-folded degeneracy of the entanglement spectra under the open(periodic) boundary condition in Fig. \ref{fig:entanglement_spectra}.
These results suggest that the spin gap obtained in the EI phase is given by the same mechanism as that of Haldane gap.
\par

\section{Summary} \label{sec:summary}
We made a ground-state phase diagram of the low-energy effective model of the one-dimensional two-orbital Hubbard model for the spin-crossover region on the basis of the density matrix renormalization group method.
We found an excitonic insulating (EI) phase and spin-state ordering (SSO) phase in the intermediate region between a low-spin (LS) phase and a high-spin (HS) phase.
The EI phase is realized in the region where $t_a/t_b\sim 1$, and the SSO phase is realized in the region where $t_a/t_b\ll 1$, where $t_a/t_b$ is the ratio of electron transfers.
\par
The spin correlation function in the EI phase has a peak wavenumber at $k=\pi\braket{P_H}$, resulting in an incommensurate spin correlation. The spin-state structure in the SSO phase shows the LS/HS/HS structure and various types of incommensurate structures depending on the crystal-field splitting.
\par
We also made a phase diagram in a magnetic field. We found a spin gap not only in the LS and HS phases but also in the EI phase.
The magnetization process in the EI phase has two stages, where the magnetization curve has different gradients corresponding to the phase transition, where the directions of $s=1$ spins are fixed by a magnetic field.
\par
In the entanglement spectra in the EI phase, the degeneracy in the largest singular values was found. The degree of degeneracy depended on the boundary conditions, which suggested the existence of the edge state similar to the valence bond solid state in the AKLT model.

\section*{Acknowledgements}
We are grateful to M. Naka, J. Nasu, A. Ono, and Y. Masaki for their fruitful discussion and critical readings of the manuscripts. H. M. also acknowledges financial supports from KAKENHI No. 21K03380, CSRN and CSIS in Tohoku University.

\bibliographystyle{jpsj.bst}
\bibliography{70347.bib}

\end{document}